# Comprehensive treatment and analysis of Fishpond sediments as a source of organic fertilizers

**Olufunke O. Oyebamiji[1], Peter O. Balogun[1], Najeem O. Oladosu[2], Stephen Emmanuel[3], Abiodun O. Adeoso[4]**

[1]Department of Biological Sciences, School of Science, Yaba College of Technology, Yaba, Lagos, Nigeria.
[2]Department of Chemistry, University of Lagos, Akoka Yaba, Lagos, Nigeria.
[3]Department of Microbiology, Kogi State University, Anyigba, Kogi State, Nigeria
[4]Nutrition & Dietetics Department, Yaba College of Technology, Lagos, Nigeria.

**Abstract:**
Agricultural fertilizers are essential to enhance proper growth and crop yield. Chemical fertilizers endanger ecosystems, soil, plants, and animal and human lives. This has increased interest in biofertilizers which are products that contain living microorganisms or natural compounds derived from organisms such as bacteria, fungi, and algae that improve soil chemical and biological properties, stimulate plant growth and restore soil fertility. This study aimed to use natural coagulants in the management of fishpond sediment for the potential recovery of nutrients that can be utilized in crop production as organic fertilizers. Fishpond wastewater and sediment were collected, treated with natural, powdered *Moringa oleifera* seed coagulant, and dried. Greenhouse evaluation of the dried sediment was carried out using maize seeds on soil primed with the sediment. Growth parameters such as plant height, leaf length, root length, plant weight, and percentage yield were determined. The nitrate, total nitrogen, phosphate and total phosphorus content of the organic fertilizer were also determined. Plant height in maize showed a 35.8 - 44.4% increase in fertilized maize and a 20.3 - 39.3 % increase in the leaf length. There was a 38.8 - 66 % increase in root length and a 23.9 – 36.5 % increase in plant weight in fertilized maize. The nutritive component was 59.80 mg/Kg, 655.56 mg/Kg, 103.87 mg/Kg and 426.60 mg/Kg respectively. For a period of 21 days of active growth, the growth rate of maize primed with organic fertilizer was 0.93 cm/day with a total yield of 100 % while the control was 0.68 cm/day with a total yield of 50 %. A dried biofertilizer that can improve plant growth, weight, and percentage yield without having to deal with smelly manure looks promising and seems to be a better alternative to explore in agriculture.

# 1.0 Introduction

Agricultural fertilizers are meant to enhance the growth and yield of crops instead, they have endangered ecosystems, the environment and living things generally leaving the soils degraded, polluted, and less productive. In the nearest future, one of our major challenges will be how to meet our food demands and still maintain a healthy environment for all living forms (Godfray *et al*., 2010; Odegard and van der Voet, 2014). A major threat resulting from global climate change is the reduction in harvests despite increased technology in many areas of the world. This poses a series of challenges to the populace which require increasing agricultural production and developing technologies to improve yield while preserving and conserving the environment (Tilman *et al*., 2002; Foley *et al.,* 2011). These are reasons why organic fertilizers/ biofertilizers are becoming the preferred option over chemical fertilizers in agriculture. Biofertilizers are sustainable, cost-effective, and environmentally friendly as they not only improve agricultural production but also enhance the soil and reduce environmental pollution (Kawalekar, 2013). Biofertilizers are made up of living microorganisms and/or metabolic compounds produced by these organisms e.g bacteria, fungi, and algae that improve the chemical and biological properties of the soil, improve plant growth, and the fertility of the soil (Abdel-Raouf *et al.,* 2012). Biofertilizers are a cheap and renewable source of nutrients that gained importance due to their low cost among peasant farmers (Ritika and Utpal, 2014). The interest is increasing, however many of the products that are available are often of very poor quality, and basic knowledge of which is best for which crop is lacking, resulting in the loss of confidence from farmers. Fishpond sediment contains lots of nutrients and organic matter and is often discharged into the environment as this is toxic to the fish. Therefore, it may have potential to serve as fertilizer in crop production. However, it contains compounds that undergo rapid degradation thereby producing unpleasant odors and becoming a threat to the environment. There is a need to manage and handle it's discharge into the environment efficiently in a sustainable manner (Danuta *et al.,* 2020). The overall goal of this study was to use natural coagulants in the management practices of fishpond sediment for the potential recovery of nutrients and its use in the crop production as organic fertilizers.

## 2. 0 MATERIALS AND METHODS

### 2.1 Fishpond wastewater collection

Fishpond wastewater samples were collected from four fishponds used in rearing catfish (*Clarias gariepinus*) located in Abule-egba area of Lagos State and Shodipe Town, Olambe of Ogun State, Nigeria. The tap of the ponds was opened to remove excess water and subsequently, the sediment mixed with wastewater was collected from the base of the fishponds using a clean plastic container into a larger plastic 25 L container.

### 2.2 Coagulant preparation

*Moringa oleifera* seed, which is a plant-based natural coagulating aid was used. Dried and matured Moringa seeds were collected and ground with a domestic blender until a fine particle was obtained. This powdered form of moringa with varying dosages was used as a coagulant for the wastewater treatment (Akinwole *et al.,* 2016; Akhila *et al.,* 2019)

### 2.3 Fishpond wastewater sedimentation

The wastewater collected in a large plastic container was seeded with powdered moringa seed. After 24 hours, the supernatant on the coagulated suspensions was decanted and the sediment was recovered using a muslin cloth filtration to obtain all the particles formed in the wastewater. The fishpond wastewater sediments were shade-dried at room temperature for 3 days and placed in paper bags.

### 2.4 Determination of organic composition and pH of the sediment

The organic matter content of the dry sediment was determined using a Thermolyne 1400 muffle furnace. This was burned at 550 °C for 2 hours to determine the quantity of organic matter (Dróżdż *et al.,* 2020). The pH determination was carried out in 0.01 M $CaCl_2$ solution at a sediment-to-solution ratio of 1:1 using a pH meter (Mettler Toledo). The pH meter was calibrated with standard buffers (Mettler Toledo) of pH 4.01, 7.00, and 9.21 prior to analysis (APHA, 2017).

## 2.5 Determination of the elemental composition of the sediment

To determine the elemental composition, 1.0 g of each dried sediment was weighed, transferred into a Kjeldahl flask and 25 mL of aqua regia (mixture of concentrated hydrochloric acid and nitric acid in ratio 3:1 respectively) was added. The mixture was digested on a hot plate at $120^0C$ in a fume cupboard with the occasional addition of aqua regia until a clear solution with less than 5 ml was obtained. The digest was poured into a 50-mL standard flask and made up to the mark with distilled deionized water. This was filtered into a clean plastic bottle using Whatman no. 540 filter paper and stored in the refrigerator prior to analysis (APHA, 2017). The digests were analyzed for elemental contents using an inductively coupled plasma-optical emission spectrometer (Agilent 5900 ICP-OES, USA). Agilent multi-element standards (500 mg/L Calibration Mix Majors, 100 mg/L Calibration Mix 1 and 100 mg/L Calibration Mix 2) were diluted with 15 % hydrochloric acid (v/v) and 5 % nitric acid (v/v) to prepare working standards. The working standards were run to obtain the calibration graph for each element.

## 2.6 Digestion for Total Nitrogen and Total Phosphorus

For the determination of total nitrogen and phosphorus, 1.0 g of dried sediment was digested using 3.6 g of $K_2S_2O_8$ in 2 mL of 0.3 M NaOH (Oladosu *et al*., 2017). The total nitrogen of the solution was obtained from the nitrate determination using cadmium reduction with Griess-Ilosvay spectrophotometric method. Total phosphorus was obtained from the phosphate determination using Murphy-Riley colorimetric method (APHA, 2017).

### 2.6.1 Nitrate Determination

Sediment (2.5 g) was extracted with 10 mL of 2 M KCl, shaken at 220 rpm using a mechanical shaker for 20 min at room temperature (Oladosu et al., 2017) to obtain nitrate solution prior to its colorimetric determination. Nitrate was determined by cadmium reduction of nitrate to nitrite using the Griess-Ilosvay method (APHA, 2017). A standard curve was prepared using the nitrate standard solutions in the range of 0.05 to 1.0 mg $L^{-1}$ $NO_3^-$ -N (Olayinka *et al*., 2016).

**2.6.2 Phosphate Determination**

Sediment (4.0 g) was extracted with 50 mL of 0.5 M $NaHCO_3$ (pH 8.5) and shaken at 220 rpm using a mechanical shaker to obtain phosphate solution prior to its colorimetric determination. The extraction temperature and time were 25°C and 20 min respectively (Oladosu *et al*., 2016). Phosphate extract was determined by the chromogenic-mineral complex formed from the oxidation-reduction reaction of Murphy and Riley's (APHA, 2017) method. A calibration curve was generated using ten orthophosphate working standard solutions in the range of 0.050 to 0.500 mg $L^{-1}$ $PO_4^{3-}$-P (Olayinka *et al*., 2016).

## 2.7    Greenhouse evaluations

Agricultural soil was mixed with dried fishpond sediments (biofertilizer) by using a ratio 3:1 that is, three bowls of soil to 1 bowl of dried fishpond sediments (biofertilizer) and this was thoroughly mixed. The soil-sediment mixture was sprinkled with water and placed in a bag perforated with holes to allow aeration and aid degradation. This combination was prepared every two weeks for six weeks to prime the soil with biofertilizers. The soil-sediment mixture and the pristine soil (control) were poured into planting trays already perforated and this was used for planting maize. Maize seeds (20) per tray were planted and watered every 3 days with 750 ml of water and growth parameters were measured (Dey and Raghuwanshi, 2020) until the experiment was terminated.

## 2.8    Growth Parameters

### 2.8.1   Seedling count

The seedling counts to determine the survival rate were carried out by counting the number of seeds that sprouted on each tray starting one week after the seeds were planted, Observations were then recorded (RA, 2020).

### 2.8.2   Plant height

The height of the plant was measured with flexible twine by holding it close to the stem of the crop. The measurement was taken from the ground level at the point on the stem where roots start to grow to the leaf base of the highest leaf. This was done at different intervals.

### 2.8.3 Leaf length

The length of leaves on each tray was also measured using the flexible twine to measure carefully from the base of the leaf attached to the stem to the tip. This was done at different intervals.

### 2.8.4 Plant weight

This was done at the end of the experimental period. The plants were gently removed from the soil and the root area was washed off with water to remove soil particles. The plants were gently blotted with a soft paper towel to remove moisture (Wood and Roper, 2000). The plant was placed on the weighing balance and the readings were taken.

### 2.8.5 Root length

The plants were carefully handled while taking the root measurement, the root was measured with tape rule and the readings were recorded.

## 3.0 RESULTS

## 3.1 Physical Properties of The Organic Fertilizer Produced

The wet fishpond wastewater sediment (sludge) had a marshy -mid texture (brown colour) before drying and a sandy texture (dark grey colour) after drying.

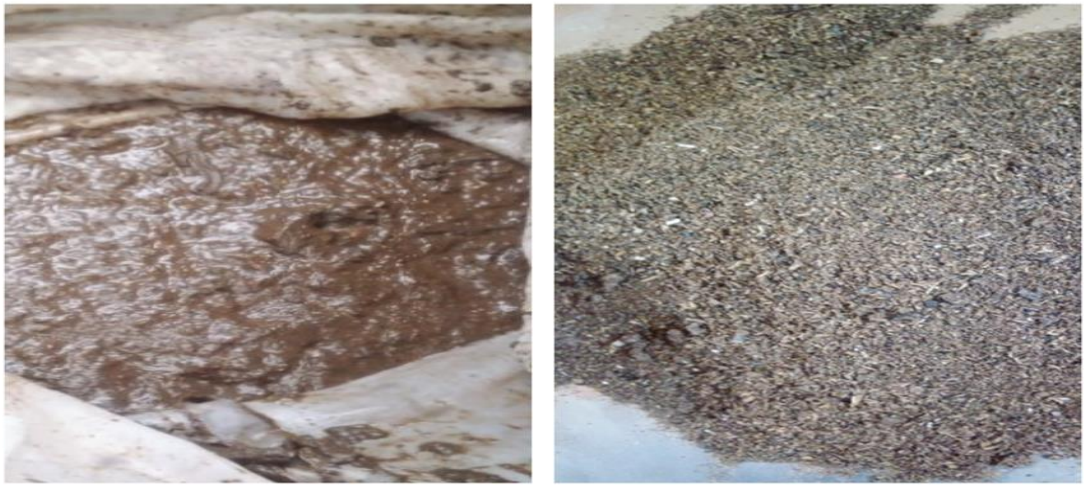

**Plate 1** The physical structure of both wet and dried coagulated fishpond wastewater sediment

## 3.2  Growth Parameters in The Five Experimental Trays.

There were noticeable variations in the maize plant parameters across the five experimental trays.

### 3.2.1 Leaf length

There was variation in the Leaf length of maize monitored for the period of 32 days. Figure 1 shows the graph comparing the average leaf length in five trays with different priming times and control. The average leaf length per tray ranged from 12.29 to 16.22 cm.

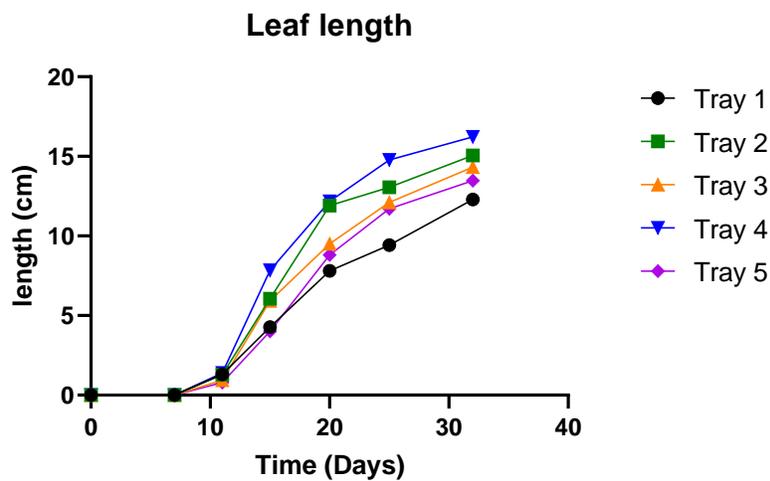

**Figure 1: The average leaf length of the maize plant (cm) for a period of 32 days across the five experimental set-ups.**

### 3.2.2  Plant height

Plant height (Figure 2) was also monitored during the period. The average plant height in the five trays with different priming times and control ranged from 16.25 to 22.07 cm.

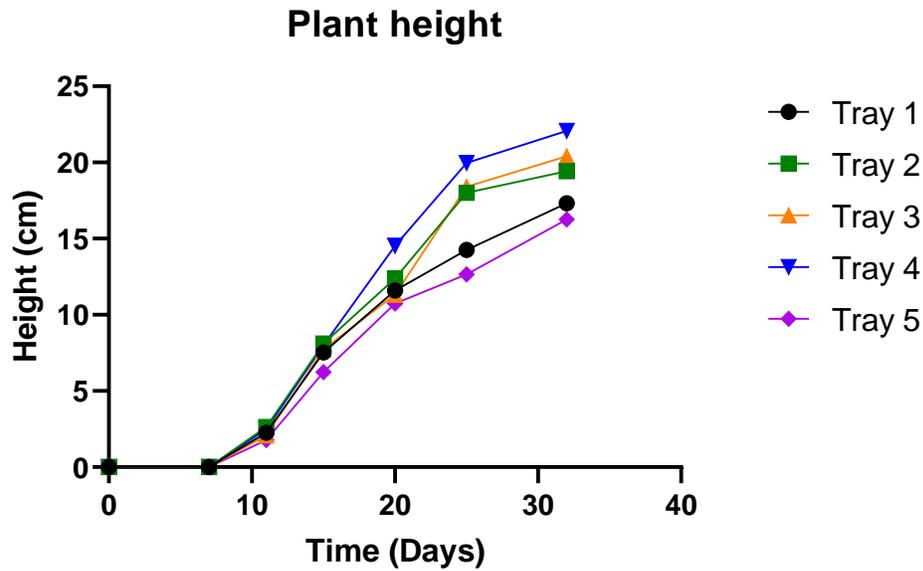

**Figure 2: The average plant height in maize plants (cm) in the five different trays for a period of 32 days across the five experimental set-ups.**

## 3.3 The Root Length, Wet Weight, and Percentage Yield

The table below (table 1) shows the percentage yield of the maize plant was between 50 % (in control) and 100 % (in tray 4). Tray 4 with the highest percentage yield is the tray that has all its seeds germinated into seedlings. This shows that priming of these organic fertilizers before planting was not necessary and if done, it could reduce the yield of maize crops. The average root length of the maize plant in each tray ranged from 3.84 - 5.33 cm, and the wet weight of the maize plant (average of 10) after 32 days ranged from 3.53 - 4.82 g.

**Table 1: The percentage yield and physical parameters of maize plant**

| Plants | Tray 1 | Tray 2 | Tray 3 | Tray 4 | Tray 5 |
|---|---|---|---|---|---|
| Percentage yield (%) | 75 | 85 | 60 | 100 | 50 |
| Root length (cm) | 4.25 | 4.37 | 4.77 | 5.33 | 3.84 |
| Wet weight (gm) | 3.83 | 3.9 | 4.4 | 4.82 | 3.53 |

## 3.4 Physico-chemical Composition of the Fishpond Sediment

The result below shows the Nutritive components of the organic fertilizer produced. pH content in the organic fertilizer is 5.03 – 5.29 and total Organic Carbon is 4.31 – 20.51 %. The phosphate content is 103.87 – 150.15 mg/kg while the Nitrate content is 59.80 – 417.08 mg/kg but the total phosphorus content and the total Nitrogen content are 426.60 – 862.07 mg/kg and 655.56 – 961.91 mg/kg respectively.

**Table 2: Nutritive composition of the fishpond sediment**

| Parameter | Batch 1 | Batch 2 |
|---|---|---|
| pH | 5.29 | 5.03 |
| Total Organic Carbon (%) | 20.51 | 4.31 |
| Phosphate (mg/kg) | 150.15 | 103.87 |
| Nitrate (mg/kg) | 417.08 | 59.80 |
| Total Phosphorus (mg/kg) | 862.07 | 426.60 |
| Total Nitrogen (mg/kg) | 961.91 | 655.56 |

## 4  DISCUSSIONS

The physical structure of the fishpond sediment looks like dark large particulate soil. According to Pike County Conservation District PCCD (2016), the dark color is an indication of increased decomposed organic matter content otherwise called humus. It is usually very fertile with a high degree of aeration. Although the pH of the fishpond wastewater ranged between 7.65 to 7.89, the pH of the organic fertilizer produced from it is acidic (5.03- 5.29). The total organic carbon ranged between 4.31 – 20.51 % and soil organic carbon is an effective energy source for microorganisms present in the soil. This improves the soil structure and crop growth (Shaji *et al*., 2020), improves soil aeration, water retention and drainage capacity, and diminishes nutrient leaching and erosion (Debska *et al*., 2016).

The phosphate/ total phosphorus content of the bio-fertilizer and the nitrate/ total nitrogen content as shown in table 2 indicates a sufficient immediate phosphate and nitrate content and high-level phosphorus and nitrogen available in organic form to be released over a period. Shaji *et al*. (2020), stated that the use of organic fertilizers will reduce the necessity of repeated application of chemical fertilizers as they slowly release nutrients into the soil and also contain many trace elements required by plants that are not contained in synthetic fertilizers.

The impact of the organic fertilizer on the maize plant is evident in the leaf length growth rate (with 0.71 cm $d^{-1}$ as against 0.60 cm $d^{-1}$ in control), plant height growth rate (with 0.93 cm $d^{-1}$ as against 0.69 cm $d^{-1}$ in control), root length (38.8 % increase), wet weight (36.5 % increase) and percentage yield (100 % as against 50 % in control) of maize plants grown with it. The obtained results showed that organic fertilizer obtained from the fishpond sediment improved the growth characteristics of the maize plant. Reports from El-Azab and El-Dewiny (2018) and Iwuagwu *et*

*al*. (2013) both showed that organic fertilizers (Biofertilizers) application improved the growth of maize roots and vegetative parts and this in turn increased the biological functions of the plant such as growth and productivity in terms of quantity.

Root length can be the controlling variable for water and nutrient uptake. As such, it was important to quantify them. According to Chen *et al*., (2020), when a plant is well nourished, the roots tend to grow longer thereby making the plant stronger at withstanding fluctuations in water and nutrient supply. They proposed that this will help the plants to be tolerant of adverse conditions so they can produce a high yield of fruits and vegetables. This was evident in the average root length of the fertilized maize plants which was 5.33 cm in tray 4 as against 3.84 cm in tray 5 which is the control. Thus, the 38.8 % increase in the root length of the maize plants gave the resultant increased plant height, increased leaf length, and overall percentage yield.

## 5 CONCLUSION

This research demonstrated the presence of organic carbon, nitrate/ total nitrogen, and phosphate/ total phosphorus in the fishpond sediments which can serve as an agricultural fertilizer supplement or soil conditioner for plant growth. This can also be applied as a waste management approach in aquaculture for environmental sustainability in a circular economy where the waste from aquaculture is pretreated to serve as the input in plant cultivation.